\documentclass[12pt]{article}

\usepackage{epsfig}
\usepackage{amsmath}
\usepackage{multicol}

\newcommand{\sty}{\scriptstyle}
\def\beq{\begin{equation}}
\def\eeq#1{\label{#1}\end{equation}}
\def\eeqn{\end{equation}}

\def\beqa{\begin{eqnarray}}
\def\eeqa#1{\label{#1}\end{eqnarray}}
\def\eeqan{\end{eqnarray}}

\let\bar=\overbar

\def\Dslash{\not{\hbox{\kern-4pt $D$}}}
\def\dslash{\not{\hbox{\kern-2pt $\del$}}}

\def\msb{{\bar{\ssstyle M \kern -1pt S}}}

\usepackage{fancyhdr,graphicx}
\def\Title#1{\begin{center} {\Large {\bf #1} } \end{center}}

\begin{document}

\Title{Luminosity function of GRBs}

\bigskip\bigskip


\begin{raggedright}

{\it Luis Juracy Rangel Lemos\index{Vader, D.}\\
Instituto Tectnol\'ogico de Aeron\'autica (ITA)\\
Pra\'ca Marechal Eduardo Gomes, 50 - Vila das Ac\'acias\\
12.228-900 – S\~ao Jos\'e dos Campos – SP - Brazil\\
\& Universidade Federal do Tocantins (UFT) - Campus Aragua\'ina\\
R. Paraguai, s/n (esquina com Urixamas) - Setor Cimba\\
77.838-824 - Aragua\'ina - TO - Brazil\\
{\tt Email: juracyrl@uft.edu.br}}\\

{\it Carlo Bianco and Remo Ruffini\\
Dipartimento di Fisica and ICRA, Sapienza Universit\`a di Roma\\
P.le Aldo Moro 5, I--00185 - Rome - Italy\\
\& ICRANet, P.zza della Repubblica 10, I--65122 - Pescara - Italy\\
{\tt Email: bianco@icra.it and ruffini@icra.it}}\\

{\it Manuel Malheiro - Instituto Tectnol\'ogico de Aeron\'autica (ITA)\\
{\tt Email: malheiro@ita.br}}
\bigskip\bigskip
\end{raggedright}

\section{Introduction}

BATSE was a detector that operated between 1991 and 2000 on board the Compton Gamma-Ray Observatory (CGRO) satellite (\cite{fish89a}, \cite{fish89b} and \cite{paciesas89}), which triggered 2704 GRBs (BATSE catalog \cite{fish94}). Due to this rich amount of data, and in spite of the lack of any redshift measurement for GRBs before the 28$^{th}$ of February 1997 \cite{costa97}, Schmidt used them to build a sample of GRBs, called by him the GUSBAD catalog \cite{sch04}, with 2207 GRBs, and to perform \cite{sch09} a detailed statistical analysis over this sample. His goal was to test the use of $V/V_{max}$ as a distance estimator, to estimate the effect of Malmquist bias, and to obtain an Amati-like relation \cite{amati02} between $E_{pk}-L_{iso}$. Schmidt in his paper \cite{sch09} used a reduced version of the GUSBAD catalog; he took out the GRBs with peak photon flux ($F_{pk}^{ph}$) less than the limiting flux as $F_{lim}=0.5\,{\rm ph\,cm^{-2}s^{-1}}$, to be sure to have a complete subsample not affected by 
selection effects. This reduced version of the GUSBAD catalog contains 1319 GRBs.

Within the Fireshell model \cite{fireshell09} it is defined a ``canonical'' GRB bolometric light curve, formed by two physically distinct components: 1) the Proper-GRB (P-GRB), which is the flash emitted when the ultrarelativistically expanding $e^+e^-$-baryon plasma originating the GRB reaches transparency; and 2) the extended afterglow, which is the prolonged emission due to the interaction of the accelerated baryons, which constituted the fireshell baryon loading and have been left over after transparency, with the CircumBurst Medium (CBM) and which comprises a rising part, a peak and a decaying phase \cite{ruffini01}. What is usually called ``prompt emission'' is actually formed by both the P-GRB and the rising part and the peak of the extended afterglow, while what is usually called ``afterglow'' is the decaying phase of the extended afterglow. The ratio between the total energies emitted in the P-GRB and in the extended afterglow is ruled by the plasma baryon loading parameter $B=M_Bc^2/E_{tot}$, where 
$M_B$ is the mass of the baryons and $E_{tot}$ is the total initial energy of the plasma. If $B < 10^{-5}$ the P-GRB is energetically dominant over the afterglow, and we have a ``genuine'' short GRB. If $B \sim 3.0\times 10^{-4}$ \cite{guga11} the extended afterglow is energetically dominant over the P-GRB and, depending on the CBM average density $n_{CBM}$, we can have a ``standard'' long duration GRB (for $n_{CBM} \sim 1$ particle/cm$^3$), where the P-GRB is at most a small precursor of the main event, or a ``disguised'' short GRB (for $n_{CBM} \sim 10^{-3}$ particles/cm$^3$, a density typical of a galactic halo environment). In particular, this last case is characterized by an extended afterglow peak which is ``deflated'' by the low CBM density and has therefore a lower peak luminosity than the P-GRB in spite of having an higher total energy. (\cite{grazia07}; \cite{caito08}; \cite{caito10}; \cite{guga11}). For $10^{-5} < B < 3.0\times 10^{-4}$ the picture depends on the value of $E_{tot}$.

Following this interpretation, the GUSBAD sample, even the reduced one, contains sources of all three above mentioned kinds: ``genuine'' short, ``disguised'' short and ``standard'' long duration GRBs. It is therefore not an homogeneous sample. Moreover, the analysis done by Schmidt implies extracting the properties of each event at the moment when it reaches its peak flux. For ``genuine'' short and ``disguised'' short GRBs this happens during the P-GRB phase, while for ``standard'' long duration GRBs it happens at the peak of the extended afterglow. Since the P-GRB and the extended afterglow are due to completely different physical processes, they cannot be analyzed together. We want therefore to repeat the Schmidt analysis on a ``sanitized'' GUSBAD sample, where we tried to eliminate all the P-GRBs building an homogeneous sample of events dominated by the extended afterglow phase. This reduced the number of events of the sample from 1319 to 888.

\section{Obtaining GUSBAD ``sanitized'', spectral classes, $E_{pk}^{rp}$ and $<V/V_{max}>$}

We obtained the GUSBAD ``sanitized'' sample from the reduced version of GUSBAD (1319 GRBs), where we took out the sources whose peak luminosity are reached during the P-GRB phase; in this case we want to study only the GRBs whose peak fluxes was located in the extended afterglow, instead of the P-GRB \cite{fireshell09}. Thus we excluded the sources with at least one of the following two characteristics: 1) with duration smaller than 2 seconds or 2) with duration greater than 10 seconds and peak flux occurring in the first two seconds. With this selection we were left 888 GRBs. We want to know how much the Schmidt results are modified using such ``sanitized'' GUSBAD sample.

We now apply on the ``sanitized'' sample procedure of Schmidt \cite{sch09}. To distribute the 888 GRBs in five spectral classes, we used the Band function \cite{band93} and the peak photon flux of each spectral channel. Each spectral channel of DISCLA BATSE detector has the following energy range: $1ch:20-50$ keV, $2ch:50-100$ keV, $3ch:100-300$ keV, $4ch:\, > 300$ keV. For the upper limit of the fourth channel, we used the indication of Kaneko et al. \cite{kaneko06}, like Schmidt \cite{sch09}, which is $600$ keV. We transformed the four spectral channels (ch) in five spectral classes (sp), as $1ch+2ch=1sp$, $3ch=2sp+3sp+4sp$ and $4ch=5sp$. The representative peak spectral energy $E_{pk}^{rp}$ of each spectral class are obtained by geometric means or statistical weight of the $E_{pk}^{rp}$ of the spectral channels. It is important to emphasize that $E_{pk}^{rp}$ is representative and not observed, because is inferred from the observation. Schmidt computed the $V/V_{ max}$ of each GRB, then the $<V/V_{ max}>$ 
of each spectral class is straightforwardly obtained. We can see in the table \ref{tab_sp} the distribution of the GRBs in each spectral class, and the values obtained to ${<V/V_{ max}>}$ and $E_{pk}^{rp}$.


\begin{table}[bht]
  \begin{center}
    \begin{tabular}{c|c|c|c|c|c|c|c}
       \hline
       sp & ch & $N_{grb}$ & $E_{pk}^{rp}$(keV) & $<V/V_{max}>$ & $log_{10}(L_c)$ & $E_0^R$(keV) & $R_0$ \\
       \hline
       1 & 1,2 & 153 & 66 & $0.452\pm {\sty 0.024}$ & 49.03 & 82 & 96.7\\
       2 & 3 & 114 & 117 & $0.407\pm {\sty 0.027}$ & 50.18 & 206 & 0.37\\
       3 & 3 & 154 & 178 & $0.312\pm {\sty 0.022}$ & 51.47 & 500 & 0.0015\\
       4 & 3 & 145 & 247 & $0.295\pm {\sty 0.023}$ & 51.68 & 725 & 0.0006\\
       5 & 4 & 322 & 420 & $0.332\pm {\sty 0.017}$ & 51.29 & 1199 & 0.005\\
       \hline
    \end{tabular}
    \caption{\small New composition of the spectral classes (sp), in which shows the number of GRBs ($N_{grb}$) where each representative peak spectral energy ($E_{pk}^{rp}$) is located. $L_c$ is the center luminosity (dimension $erg/s$), $E_0^R$ break spectral energy in rest frame and $R_0$ the GRB rate density at $z=0$ (dimension $Gpc^{-3}yr^{-1}$).}
    \label{tab_sp}
  \end{center}
\end{table}

\section{Luminosity Function, Source Count and GRB rate}

\begin{figure}[htb]
 \centering
   \epsfig{file=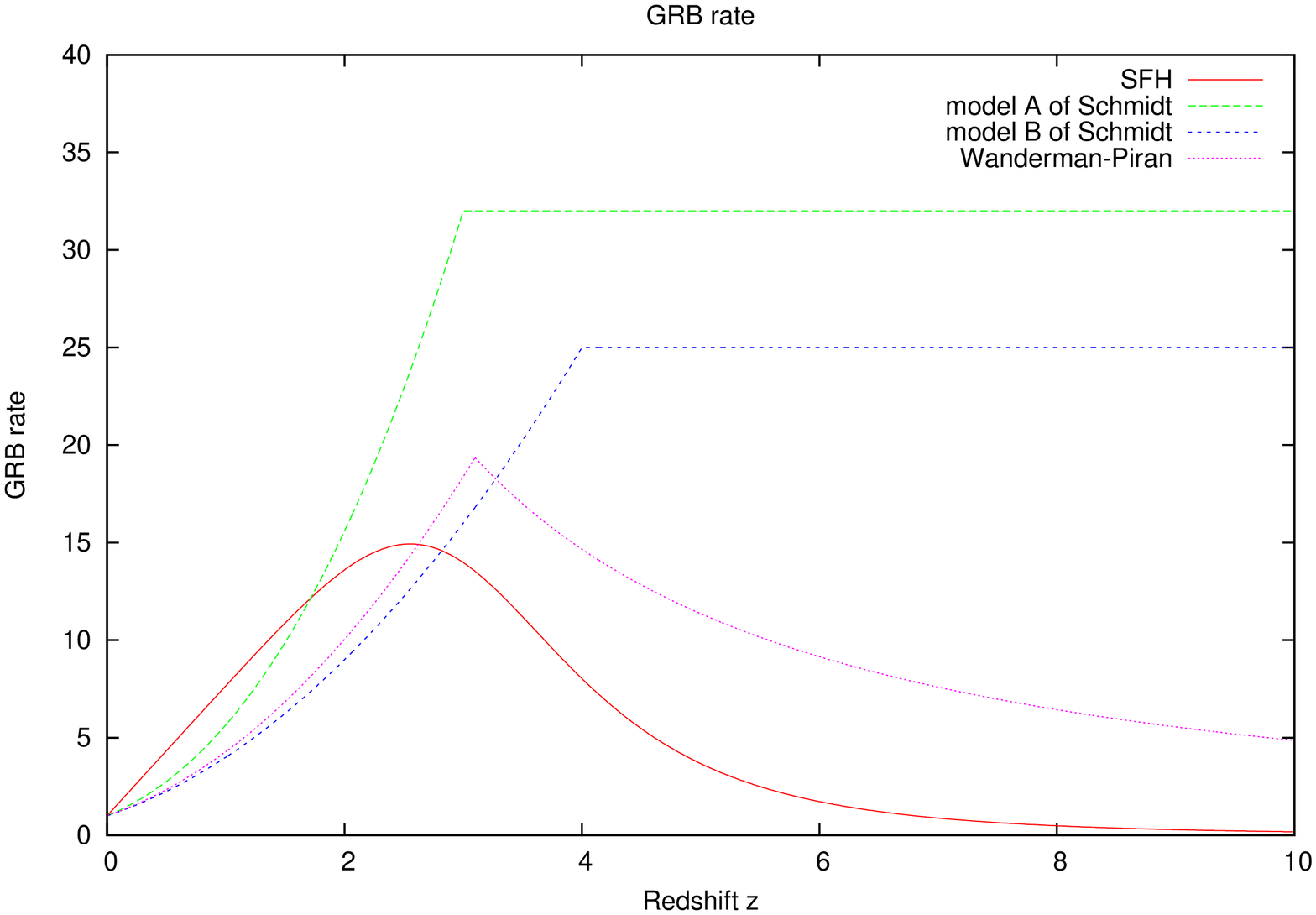,angle=-90,scale=0.27}
   \epsfig{file=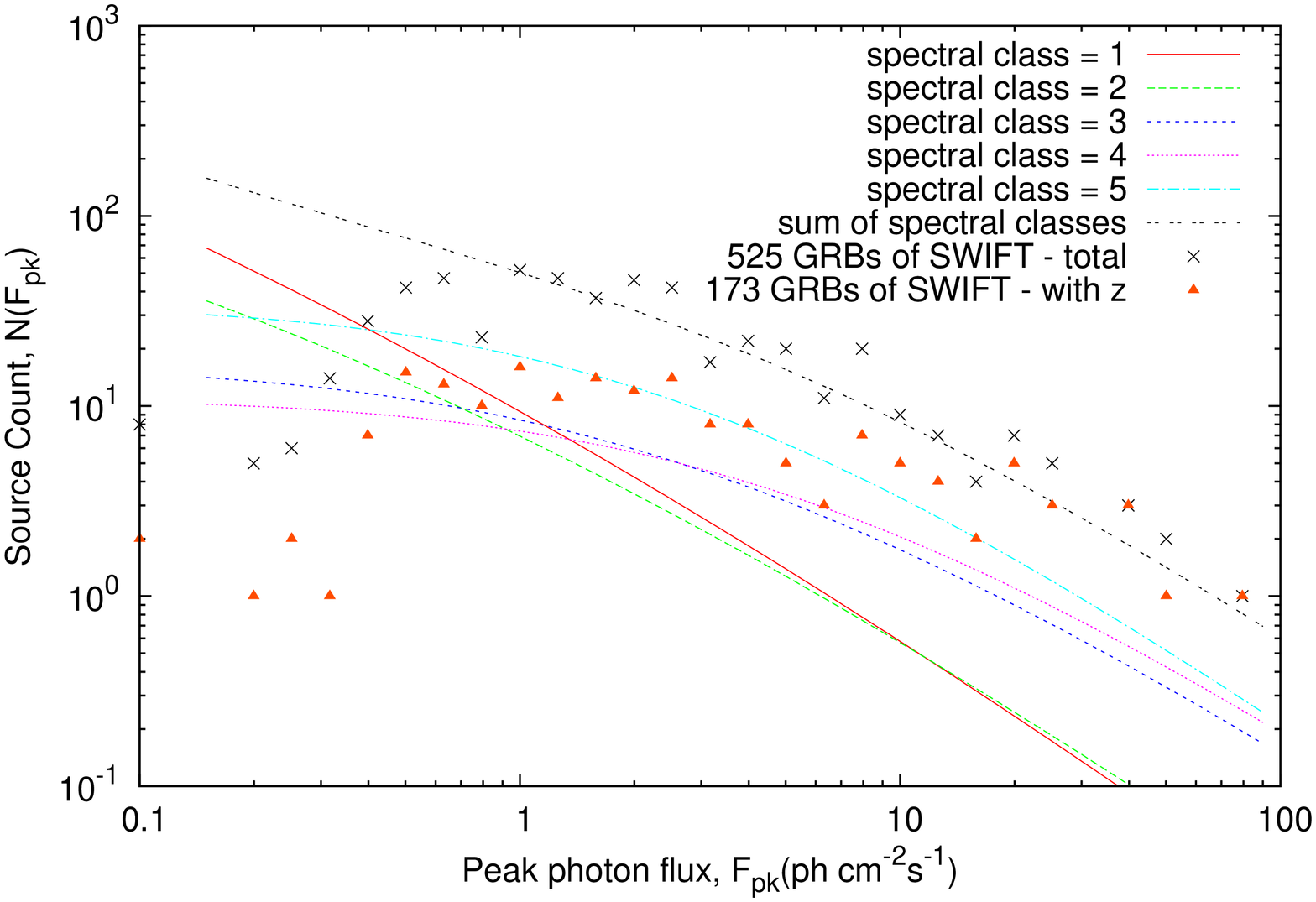,angle=-90,scale=0.27}
 \caption{\small The figure A shows some observed GRB rate densities used by the literature. The figure B shows the distribution of the peak photon flux, where the dots are from SWIFT data and the lines are the prediction of Eq. (\ref{abacaxi-3}).}
 \label{fig_GReNF}
\end{figure}


We used the luminosity function (LF) obtained by Schmidt \cite{sch09}, $\Phi(L,sp,z) = \Phi_0(L,sp)GR^R(z)$, where $GR^R(z)=GR^{ob}(z)/(1+z)$ is the intrinsic GRB rate density, and $GR^{ob}(z)$ represents the observed one in which many distant sources are missing due to their weak luminosity. Schmidt (2009) assumed five Gaussian shapes (to redshift constant) respective to the five spectral classes. The real LF in $z=0$ is given as
\begin{equation}
  \Phi_0(L,sp) = \frac{R_0(sp)}{\sigma_{\log{L}}\sqrt{2\pi}}
  \exp\left\{-\frac{1}{2}{\left[\log\left(\frac{L}{L_c(sp)}\right)/\sigma_{\log{L}}\right]^2}\right\},\label{abacaxi-2} 
\end{equation}
where $L$ is the peak luminosity, $sp$ the spectral parameter, $R_0 (sp)$ the GRB rate density at $z=0$ and $\sigma_{log{L}}$ the dispersion (or standard deviation) around the center peak luminosity $L_c(sp)$. The peak flux distribution (source count) is given by the integral of the LF, as
\begin{equation}
  N(F>F_{lim},sp) = \int_{0}^{\infty}\Phi_0(L,sp)dL\int_0^{z(L,F,sp)}\frac{GR^{ob}(z')}{1+z'}\,\frac{d{V_{com}(z')}}{dz'}dz',\label{abacaxi-3} 
\end{equation}
where $V_{com}(z)$ is the comoving volume. The upper limit $z(L,F,sp)$ is obtained from the follow constraint
\begin{equation}
  F(z,L) = \frac{L\,k(z,sp)}{4\pi\left[D_L(z)\right]^2},\label{abacaxi-4}
\end{equation}
where $k(z,sp)$ is the k-correction \cite{sch09} and $D_L(z)$ the luminosity distance.


In our approach we use the observed GRB rate density normalized at $z=0$ obtained by Wanderman \& Piran (2010) \cite{wander10}, as
\begin{equation}
  GR_{WP}^{ob}(z) = 
  \begin{cases}
    (1+z)^{2.1} &\rm{if}\, z\leq z_c\\
    c_x(1+z)^{-1.4} &\rm{if}\, z>z_c                                 
  \end{cases},\label{abacaxi-1} 
\end{equation}
where $z_c=3.1$ and $c_x=139.55$. The figure \ref{fig_GReNF}A shows the comparison among the GRB rate density of: Star Formation History (SFH), Models A and B of Schmidt (2009) \cite{sch09} and Wanderman \& Piran (2010) \cite{wander10}.

\section{Calibration of the source count}

\begin{figure}[thb]
 \centering
   \epsfig{file=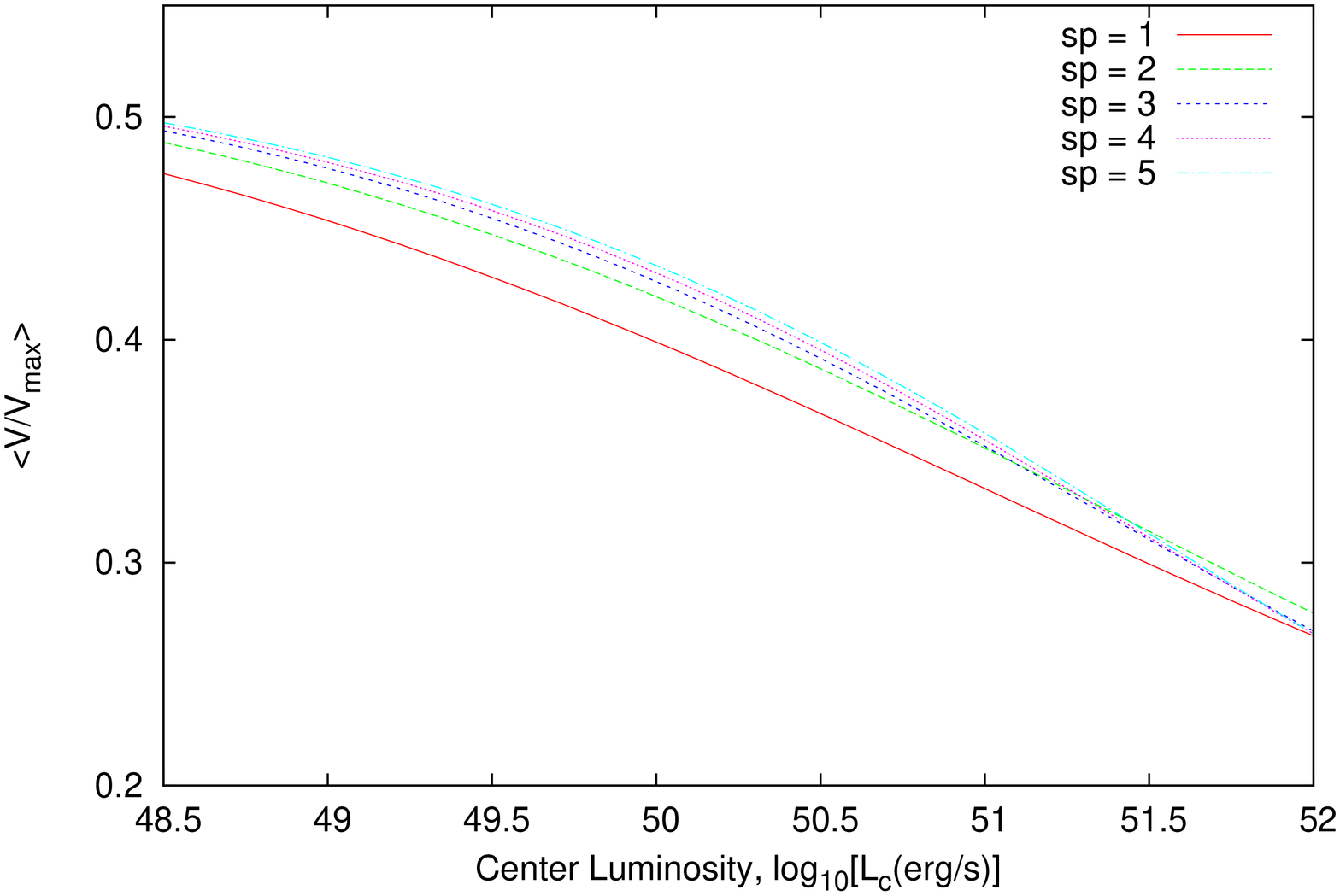,angle=-90,scale=0.27}
   \epsfig{file=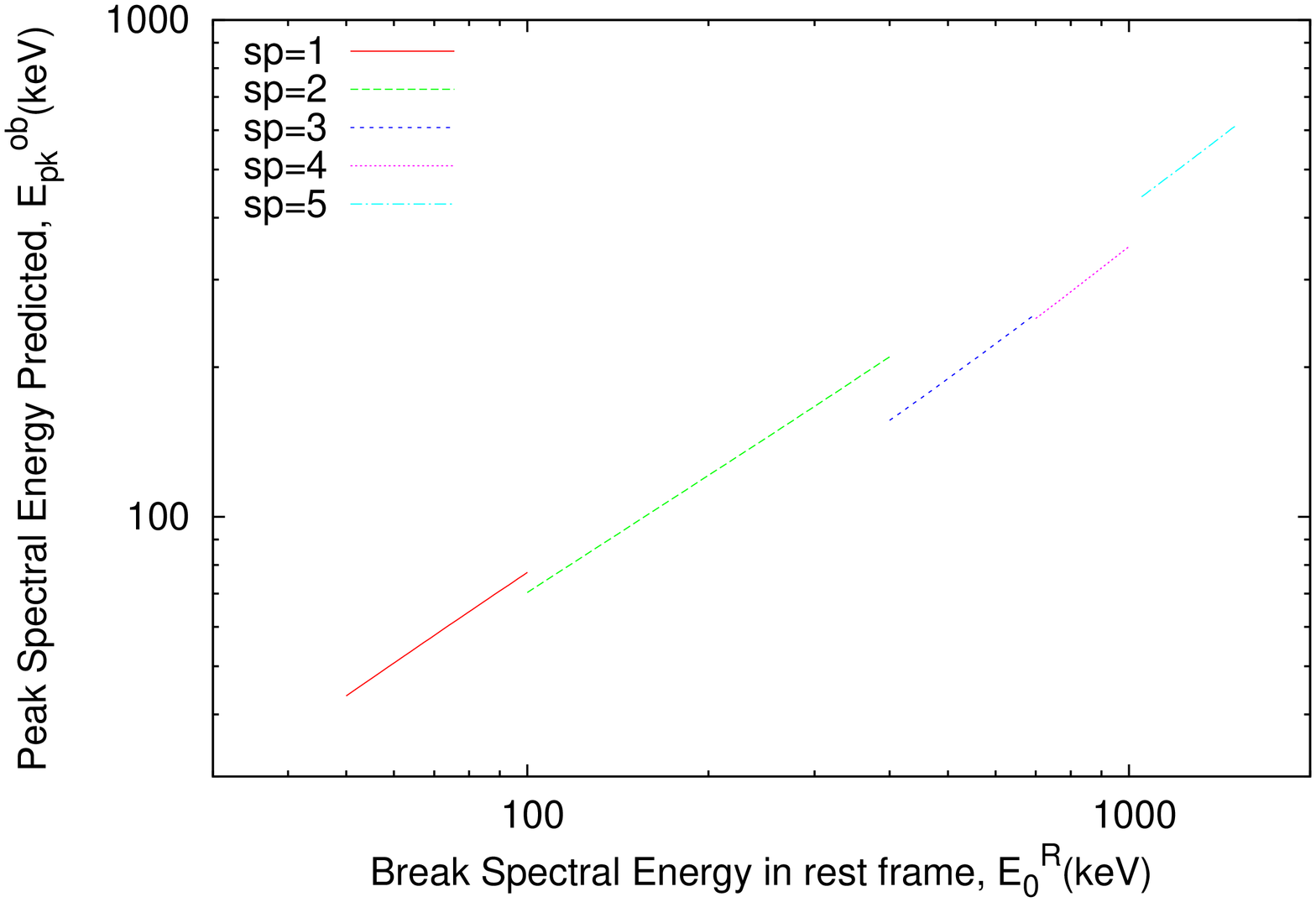,angle=-90,scale=0.27}
 \caption{\small The figures show the iterations of $L_c$ (figure A) and $E_0^R$ (figure B), obtaining in each step the predicted $<V/V_{max}>$ and $<E_{pk}^{ob}>$, respectively.}
 \label{fig_VVeEpk}
\end{figure}

To obtain the source count in function of $F_{pk}$, $z$ and $L$, before we need to calibrate three parameters of the last section: ``$L_c$'' center peak luminosity, ``$E_0^R$'' break spectral energy in rest frame and ``$R_0$'' GRB rate density at $z=0$; and the parameterization is by trial and error. We made the iteration (showed in the figures \ref{fig_VVeEpk}A and \ref{fig_VVeEpk}B) of $L_c$ and $E_0^R$ to obtain in each step $<V/V_{max}>$ and $<E_{pk}^{ob}>$, respectively; and this procedure was made until reach the $<V/V_{max}>$ and $E_{pk}^{rp}$ obtained from the ``sanitized'' sample of GUSBAD (table \ref{tab_sp}).

\section{Conclusions}

\begin{figure}[hbt]
 \centering
   \epsfig{file=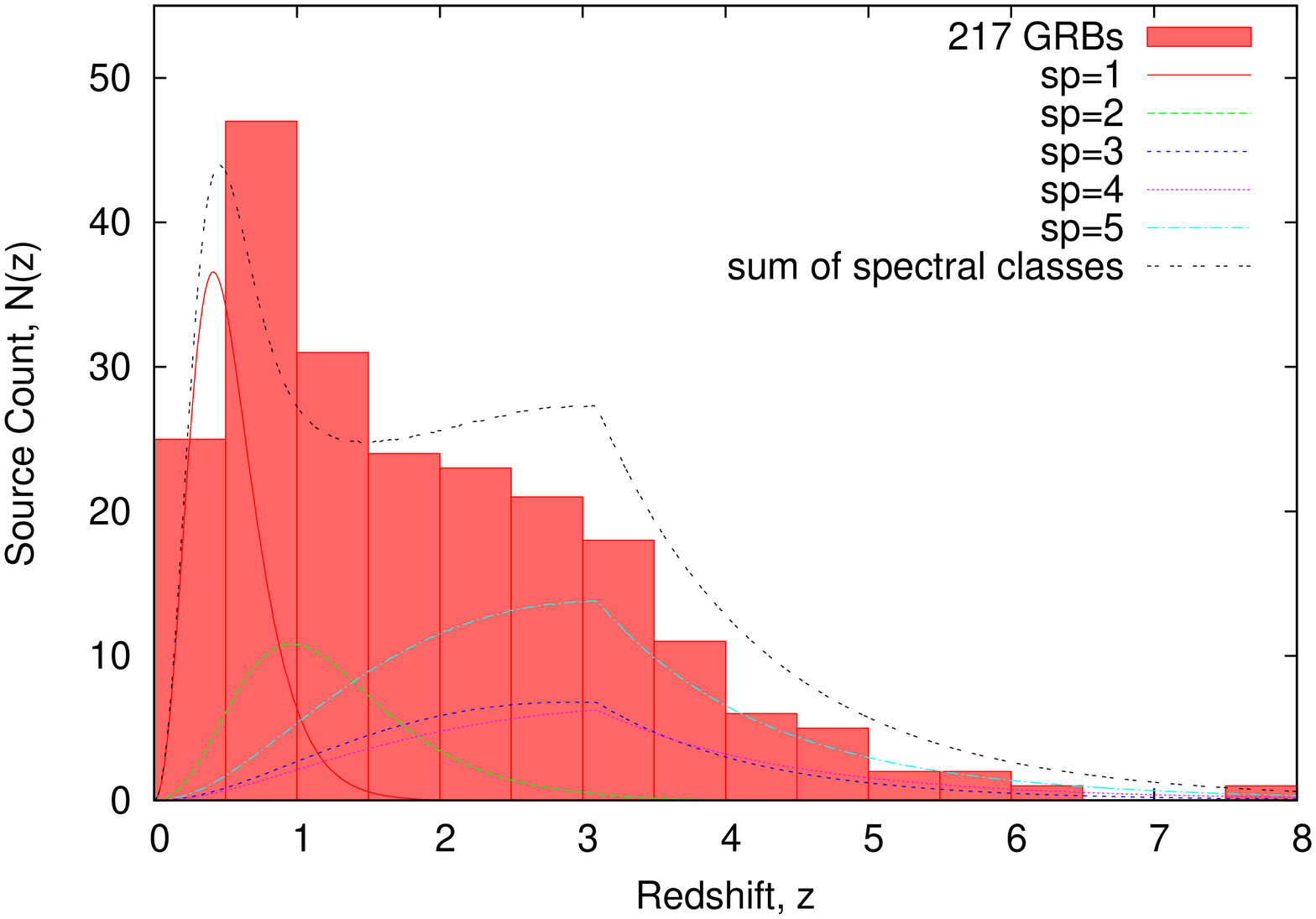,angle=-90,scale=0.27}
   \epsfig{file=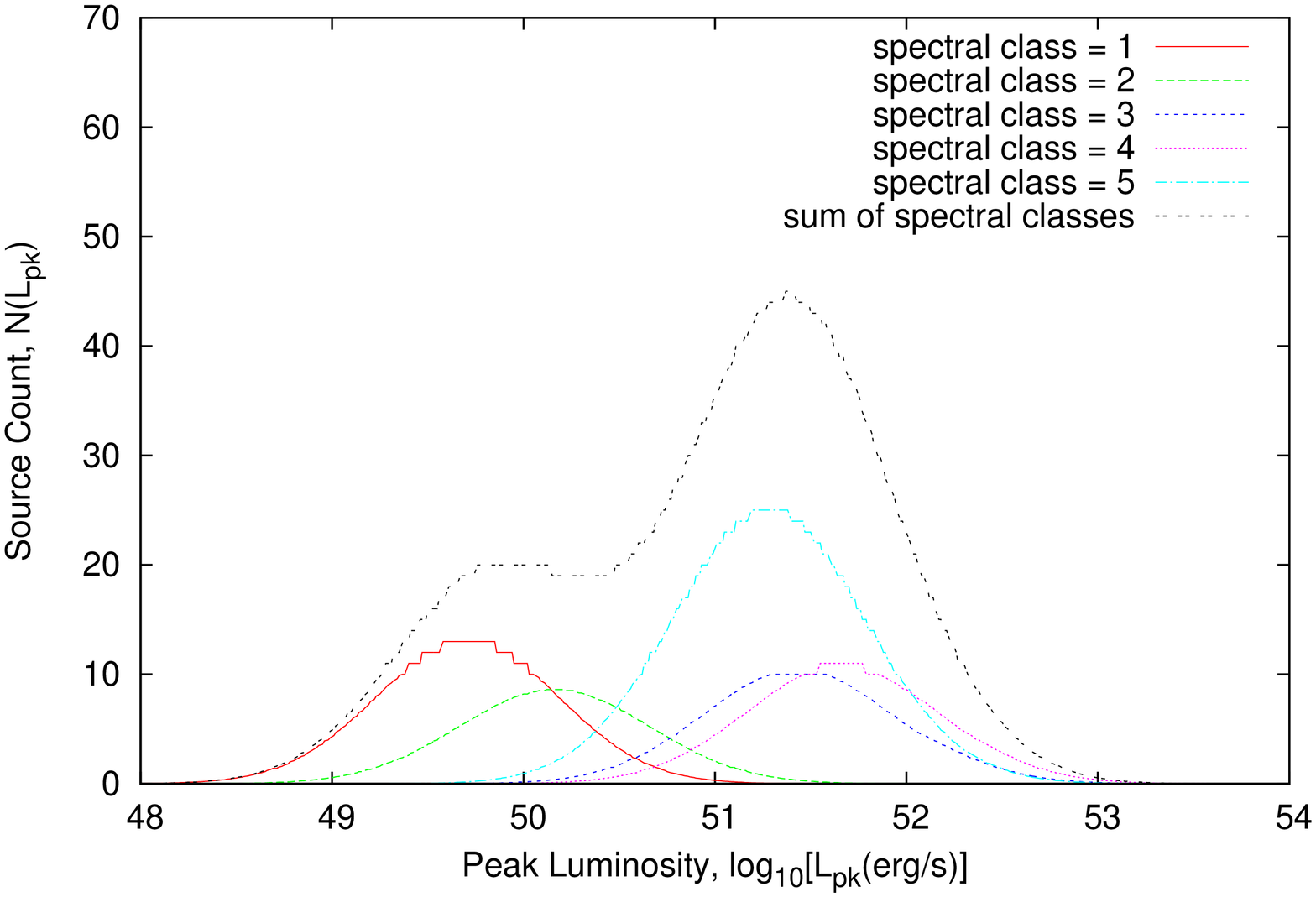,angle=-90,scale=0.27}
 \caption{\small Source count in function of the redshift (figure A) and luminosity (figure B). The curves are the predictions by Eq. (\ref{abacaxi-3}). The histogram shows the distribution of almost all GRBs with known redshift.}
 \label{fig_NZeNL}
\end{figure}

The goal in this work was to apply the statistical approach of Schimdt (2009) to a GRB sample without contamination by the P-GRBs. We have also used a different GRB rate density from the one used by Schmidt (2009), namely that obtained by Wanderman \& Piran (2010). We can see the effect of this choice in the statistics, looking at the figures \ref{fig_GReNF}B and \ref{fig_NZeNL}A, where a reasonable agreement with the observation data is obtained. The next goal is to obtain an Amati-like relation \cite{amati02}, between the isotropic luminosity and peak spectral energy in the rest frame.

\end{document}